\documentclass[a4paper]{article}

\usepackage[english]{babel}
\usepackage[utf8x]{inputenc}
\usepackage[T1]{fontenc}

\usepackage{amsfonts,amstext,amsmath,amssymb}

\usepackage[table]{xcolor}
\usepackage{listings}
\usepackage{graphicx}
\usepackage{rotating}
\usepackage{multirow}
\usepackage[toc,page]{appendix}
\usepackage{tikz}
\usetikzlibrary{arrows}
\usetikzlibrary{decorations.markings}
\usepackage{flexisym}

\usepackage{array}
\usepackage{float}

\DeclareMathOperator{\logis}{logis}
\newcommand{\R}{\mathbb{R}}

\newcommand{\Rp}{\mathbb{R}^p}

\newcommand{\calU}{{\cal U}}
\newcommand{\calW}{{\cal W}}

\newcommand{\calZ}{{\cal Z}}

\newcommand{\un}{\mathbf{1}}

\newcommand{\br}{\mathbf{r}}

\newcommand{\bt}{\mathbf{t}}
\newcommand{\bu}{\mathbf{u}}

\newcommand{\bw}{\mathbf{w}}
\newcommand{\bx}{\mathbf{x}}

\newcommand{\bX}{\mathbf{X}}

\newcommand{\by}{\mathbf{y}}

\newcommand{\bY}{\mathbf{Y}}

\newcommand{\bz}{\mathbf{z}}

\newcommand{\bbeta}{\boldsymbol{\beta}}

\newcommand{\bmu}{\boldsymbol{\mu}}

\newcommand{\bpi}{\boldsymbol{\pi}}

\newcommand{\brho}{\boldsymbol{\rho}}

\newcommand{\bSigma}{\boldsymbol{\Sigma}}
\newcommand{\btheta}{\boldsymbol{\theta}}

\newcommand{\argmax}{\mathop{\mathrm{argmax}}}

\newcommand{\diag}{\mathop{\mathrm{diag}}}

\newcommand{\Econd}[2]{\mathbb{E}\left[#1\left|#2\right.\right]}

\newtheorem{lemma}{Lemma}

\usepackage[a4paper,top=3cm,bottom=2cm,left=3cm,right=3cm,marginparwidth=1.75cm]{geometry}

\usepackage{amsmath}
\usepackage{graphicx}
\usepackage[colorinlistoftodos]{todonotes}
\usepackage[colorlinks=true, allcolors=blue]{hyperref}
\usepackage[affil-it]{authblk}

\title{Block clustering of  Binary Data with Gaussian Co-variables}
\author{Serge Iovleff$^{1}$ ,  Seydou Nourou Syllla$^{2}$  and Cheikh Loucoubar$^{2}$,\\
$^{1}$ University of Lille, France, serge.iovleff@univ-lille.f\\
$^{2}$ G4-Bio-Informatique,Bio-mathematique et Modelisation- Institut Pasteur, Dakar, Senegal, seydou.sylla@pasteur.sn, cheikh.loucoubar@pasteur.sn\\

}

\begin{document}

\maketitle

\begin{abstract}
The simultaneous grouping of rows and columns is an important technique that is increasingly
used in large-scale data analysis. In this paper, we present a novel co-clustering method
using co-variables in its construction. It is based on a latent block model taking into
account the problem of grouping variables and clustering individuals by integrating
information given by  sets of co-variables. Numerical experiments on simulated data sets
and an application on real genetic data highlight the interest of this approach.
\end{abstract}

\section{Introduction}

Classification is a method of data analysis that aims to group together a set of observations
into homogeneous classes. It plays an increasingly important role in many scientific and
technical fields. Its aim is the automatic resolution of problems by decision-making based
on the observations and to define the rules for classifying objects depending on qualitative
or quantitative variables.

Clustering is the most popular technique for data analysis in many disciplines.
In recent years, co-clustering has been increasingly used.
Unlike classical clustering, which groups similar objects from a single collection of
objects, co-clustering or bi-clustering \cite{madeira2004biclustering} aims at
simultaneously grouping objects from two disjoint sets, thus revealing interactions
between elements of two sets. 

It is most often used with bipartite spectral graphing partitioning methods in the
field of extracting text data \cite{Dhillon01} by simultaneously grouping documents
and content (words) and analyzing huge corpora unlabeled documents \cite{xu2010co}
to simultaneously understand aggregates of subsets of web users sessions and
information from the page views. Co-clustering algorithms have also been developed
for computer vision applications. It is used for grouping images simultaneously
with their low-level visual characteristics and for content-based search
\cite{guan2005spectral}.

In this paper we extend co-clustering methods allowing simultaneous detection of
associations between  variables and individuals by taking into account co-variables.
These co-variables can be additional measures of interest.
Consideration of a co-variable is expected to provide better separation of groups of
variables and especially groups of individuals. Classification quality is determined
by general validation measures specific to the co-clustering method.
This approach can be useful when co-clustering a set $\bX$ of variables and individuals
in coherence with an independent $\bY$ variable measured on these same individuals. For
example, in the co-clustering of several SNP (Single-Nucleotide Polymorphism) variables
on different patients with respect to a measured phenotype (see application in
section \ref{sec:Applications}).

The paper is organized as follows. In the first part, we explain  the principle of
block mixture models through section \ref{sec:Block mixture models}.
The latent block model for binary variable takes into account  co-variables and
the model parameters estimation  is proposed in Section \ref{sec:Latent Block model}.
The parameter estimation method is described in section
\ref{sec:Model parameters estimation}.
The choice of the optimal number of blocks and the measure of influence of each
variable on the co-variable $\bY$ is presented in the second part
(section \ref{sec:Selecting the number of blocks} and \ref{sec:Measuring Influence of
a Variable}). The method is illustrated on simulated  and real genetic data in
the last part (section \ref{sec:Applications}).

\section{Block mixture models}
\label{sec:Block mixture models}
\subsection{Classical latent block model}
Let $\bx$ be a data set doubly indexed by a set $I$ with $n$ elements
(individuals) and a set $J$ with $m$ elements (variables). We represent a
partition of $I$ into $g$ clusters by $\bz=(z_{11},\ldots,z_{ng})$ with
$z_{ik}=1$ if $i$ belongs to cluster $k$ and $z_{ik}=0$ otherwise, $z_i=k$
if $z_{ik}=1$ and we denote by $z_{.k}=\sum_i z_{ik}$ the cardinality of
row cluster $k$. Similarly, we represent a partition of $J$ into $d$ clusters
by $\bw=(w_{11},\ldots,w_{md})$ with $w_{j\ell}=1$ if $j$ belongs to cluster
$\ell$ and $w_{j\ell}=0$ otherwise, $w_j=\ell$ if $w_{j\ell}=1$ and we denote
$w_{.\ell}=\sum_j w_{j\ell}$ the cardinality of column cluster $\ell$.

The block mixture model formulation is  defined in \cite{Govaert2003} and
\cite{bhatia2017blockcluster} (among others) by the following probability
density function
$$f(\bx;\btheta)=\sum_{\bu \in \calU} p(\bu;\btheta) f(\bx|\bu;\btheta)$$
where $\calU$ denotes the set of all possible labels of $I\times J$
and $\btheta$ contains all the unknown parameters of this model. By
restricting this model to a set of labels of $I\times J$ defined by a
product of labels of $I$ and $J$, and further assuming that the
labels of $I$ and $J$ are independent of each other, one obtain the
decomposition
\begin{equation}
  \label{eq:latent_block_model}
  f(\bx;\btheta)=\sum_{(\bz,\bw) \in \cal{Z \times W}}
  p(\bz;\btheta) p(\bw;\btheta) f(\bx|\bz,\bw;\btheta)
\end{equation}
where $\calZ$ and $\calW$ denote the sets of all possible labellings $\bz$
of $I$ and $\bw$ of $J$. Equation (\ref{eq:latent_block_model}) define a
\emph{Latent Block Model} (LBM).

\subsection{LBM for binary variables with co-variables: General formulation}
\label{sec:Latent Block model}
From now, we assume that $\bx$ is a binary data set. Let $\by$ represents
a data-set (co-variables) of $\R^p$ indexed by $I$.
In order to take into account this set of co-variables the classical
block model formulation is extended to propose a block mixture model
defined by the following probability density function
\begin{equation}
\label{eq:co_latent_block_model}
 f(\bx,\by;\btheta)=\sum_{(\bz,\bw) \in \cal{Z \times W}}p(\bz;\btheta) p(\bw;\btheta)
  f(\bx| \by,\bz,\bw;\btheta)  f(\by| \bz;\btheta).
\end{equation}
By extending the latent class principle of local independence to our block model,
each data pair $(x_{ij},\by_i )$ will be independent once $z_i$ and $w_j$ are fixed.
Hence we have
$$f(\bx,\by|\bz,\bw;\btheta)=\prod_{i,j} f(x_{ij}, \by_{i}|\bz_i,\bw_j;\btheta).$$
We choose to model the dependency between $x_{ij}$ and $\by_i$ using the canonical
link for binary response data
\begin{equation}\label{eq:link}
  f(x_{ij}|\by_i,\bbeta_{z_iw_j})= \logis(\beta_{0,z_iw_j}+\bbeta_{z_iw_j}^T\by_i)^{x_{ij}}
\, \left(1-\logis(\beta_{0,z_iw_j}+\bbeta_{z_iw_j}^T\by_i)\right)^{1-{x_{ij}}}
\end{equation}
with $(\beta_0,\bbeta_{k,l})\in\R^{p+1}$ and $\logis(x) = e^x/(1+e^x)$.
Each data point $\by_i$ will be independent once $z_i$ are fixed.
In the examples presented in section \ref{sec:Applications}, we choose 
$$
f(\by|\bz;\btheta )=  \prod_{i} \phi(\by_i;\bmu_{z_i},\bSigma_{z_{i}})
$$
with $\phi$ denoting the multivariate Gaussian density in $\Rp$.

In order to simplify the notation, we add a constant coordinate $1$ to vectors $\by_i$
and write $\bbeta_{k,l}$ in the latter rather than $(\beta_{0,k,l},\bbeta_{k,l})$.

The parameters are thus $\btheta=(\bpi,\brho,\bbeta,\bmu,\bSigma)$, where
$\bpi=(\pi_1,\ldots,\pi_g)$, $\brho=(\rho_1,\ldots,\rho_d)$ are the vectors
of probabilities $\pi_k$ and $\rho_\ell$ that a row and a column belong to
the $k$th row component and to the $\ell$th column component respectively,
$\bbeta=(\bbeta_{kl})$ are the coefficients of the logistic function,
$\bmu$ and $\bSigma$ are the means and variances of the
Gaussian density. In summary, we obtain the latent block mixture model
with pdf
\begin{equation}
  \label{eq:latent_block_model_1}
  f(\bx,\by|\btheta)=\sum_{(\bz,\bw) \in \calZ \times \calW}
  \prod_{i,j} \pi_{z_i} \rho_{w_j} \logis(\by_i^T\bbeta_{z_i w_j})^{x_{ij}}
  \left(1-\logis(\by_i^T\bbeta_{z_i w_j})\right)^{1-x_{ij}}
  \phi(\by_i;\bmu_{z_{i}},\bSigma_{z_{i}}).
\end{equation}

Using the  above expression, the randomized data generation process can be described by the four steps row labellings (R), column labellings (C), co-variable data generation (Y) and data generation (X) as follows:
\begin{enumerate}
\item[(R)] Generate the labellings $\bz=(z_1,\ldots,z_n)$ according to the
  distribution $\bpi=(\pi_1,\ldots,\pi_g)$.
\item[(C)] Generate the labellings $\bw=(w_1,\ldots,w_m)$ according to the
distribution $\brho=(\rho_1,\ldots,\rho_d)$.
\item[(Y)] Generate for $i=1,...,n$  vector $\by_{i}$ according to
the Gaussian distribution $\mathcal{N}_p(\bmu_{z_{i}},\bSigma_{z_{i}})$.
\item[(X)] Generate for $i=1,...,n$ and $j=1,...,m$ a value $x_{ij}$ according
to the Bernoulli distribution $f(x_{ij}|\by_i;\bbeta_{z_i w_j})$
given in (\ref{eq:link}).
\end{enumerate}

\subsection{Model Parameters Estimation}
\label{sec:Model parameters estimation}
The complete data is represented as a vector $(\bx,\by,\bz,\bw)$
where unobservable vectors $\bz$ and $\bw$ are the labels. The log-likelihood
to maximize is
\begin{equation}
  \label{eq:loglikelihood}
  l(\btheta)=\log f(\bx,\by;\btheta)
\end{equation}
and the double missing data structure, namely $\bz$ and $\bw$, makes statistical
inference more difficult than usual. More precisely, if we try to use an EM algorithm
as in standard mixture model \cite{Dempster} the complete data log-likelihood
is found to be
\begin{equation}\label{eq:Lc_bloc_latent}
 L_C(\bz,\bw,\btheta)=\sum_k z_{.k} \log \pi_k + \sum_\ell w_{.\ell} \log \rho_\ell
  + \sum_{i,j,k,\ell} z_{ik}w_{j\ell}\log f(x_{ij},\by_i;\btheta_{k\ell}).
\end{equation}
The EM algorithm maximizes the log-likelihood $l(\btheta)$ iteratively by maximizing
the conditional expectation $Q(\btheta,\btheta^{(c)})$ of the complete data
log-likelihood given a previous current estimate
$\btheta^{(c)}$ and $(\bx,\by)$:
\begin{equation*}
  Q(\btheta,\btheta^{(c)})
  =\Econd{L_C(\bz,\bw,\theta)}{{\bx,\by,\btheta^{(c)}}}
  = \sum_{i,k}  t_{ik}^{(c)} \log \pi_k
  + \sum_{j,\ell} r_{j\ell}^{(c)} \log \rho_\ell
  + \sum_{i,j,k,\ell}  e_{ikj\ell}^{(c)} \log f(x_{ij},\by_i;\btheta_{k\ell})
\end{equation*}
where
\begin{equation*}
  t_{ik}^{(c)}=P(z_{ik}=1|\bx,\by,\btheta^{(c)}), \qquad
  r_{jl}^{(c)}=P(w_{j\ell}=1|\bx,\by,\btheta^{(c)}), \qquad
  e_{ikj\ell}^{(c)} = P(z_{ik}w_{j\ell}=1|\bx,\by, \btheta^{(c)})
\end{equation*}
Unfortunately, difficulties arise due to the dependence structure in the model,
in particular to determine $e_{ikj\ell}^{(c)}$. The assumed independence
of $\bz$ and $\bw$ in (\ref{eq:latent_block_model}) is not conserved by
the posterior probability.

To solve this problem an approximate solution is proposed in \cite{Govaert2003}
using the \cite{hathaway_86b} and \cite{neal_98} interpretation of the VEM algorithm.
Consider a family of probability distribution $q(z_{ik},w_{j\ell})$ verifying
$q(z_{ik},w_{j\ell})>0$ and the relation $q(z_{ik},w_{j\ell}) =q(z_{ik}) q(w_{j\ell})$,
for all $i,j,k,l$. Set $t_{ik} =q(z_{ik})$ and $r_{jl} = q(w_{j\ell})$,
$\bt=(t_{ik})_{ik}$ for $i=1,\ldots,n$, $k=1,\ldots,g$ and
$\br=(r_{jl})_{jl}$ for $j=1,\ldots,m$ and $l=1,\ldots,d$.
One shows easily that
\begin{equation}
 l(\btheta) = \tilde{F}_C(\bt,\br;\btheta) +
 KL(q(\bz,\bw)\parallel p(\bz,\bw|\bx,\by, \btheta))
\end{equation}
with $KL(q\parallel p)$ denoting the Kullback-Liebler divergence of distribution $p$ and $q$,
\begin{equation} \label{eq:fuzzycriterion}
\tilde{F}_C(\bt,\br;\btheta)=\sum_k t_{.k}\log \pi_k + \sum_{\ell} r_{.\ell} \log\rho_l
    +\sum_{i,j,k,\ell} t_{ik}r_{j\ell}\log f(x_{ij},\by_i;\btheta_{k\ell})  + H(\bt) + H(\br)
\end{equation}
and $H(\bt)$, $H(\br)$ denoting the entropy of $\bt$ and $\br$, i.e.
\begin{equation*}
H(\bt) = \sum_{ik} t_{ik}\log t_{ik},\qquad H(\br) = \sum_{jl} r_{jl}\log r_{jl}  .
\end{equation*}
$ \tilde{F}_C$ is called the free energy or the fuzzy criterion.
As the Kullback-Liebler divergence is always positive, the fuzzy criterion
is a lower bound of the log-likelihood and is used as a replacement for it.
Doing that, the maximization of the likelihood  $l(\btheta)$ is replaced by
the following problem
$$
\argmax_{\bt,\br,\btheta} \tilde{F}_C(\bt,\br,\btheta).
$$
This maximization can be achieved using the BEM algorithm detailed hereafter.

\subsection{Block expectation maximization (BEM) Algorithm}
The fuzzy clustering criterion given in (\ref{eq:fuzzycriterion})
can be maximized using a variational EM algorithm (VEM).
We here outline the various expressions evaluated during E and M steps.
\paragraph{E-Step:} we compute either the values of $\bt$ (respectively $\br$)
with $\br$ (respectively $\bt$) and $\btheta$ fixed (formulas (\ref{eq:RowEStep}),
(\ref{eq:ColEStep}) hereafter). Details are given in appendix \ref{app:EStep}.
\paragraph{M-Step:} we calculate row proportions $\bpi$ and column proportions $\brho$.
The maximization of $\tilde{F}_C$ w.r.t. $\bpi$, and w.r.t $\brho$, is obtained by maximizing
$\sum_k t_{.k}\log \pi_k $, and $\sum_\ell r_{.\ell}\log \rho_\ell$ respectively, which leads to
\begin{equation}\label{eq:prop-estimate}
\pi_k=\frac{t_{.k}}{n} \quad \textrm{ and } \quad \rho_\ell=\frac{r_{.\ell}}{m}.
\end{equation}
Also, for $\bt$, $\br$ fixed, the estimate of model parameters $\bbeta$ will be obtained by
maximizing
\begin{equation}\label{eq:betakl-estimate}
\bbeta_{kl}= \argmax_{\bbeta}\sum_{ij}t_{ik}r_{jl}\log f(x_{ij}|\by_i;\bbeta),\quad k=1,\ldots,g,\quad l=1,\ldots,d.
\end{equation}
Detail are given in appendix \ref{app:MStep}. Finally parameters of the Gaussian
density are given by the usual formulas
\begin{equation}\label{eq:gauss-estimate}
\bmu_k = \frac{1}{t_{.k}} \sum_{i} t_{ik} \by_i
\qquad \mbox{ and } \qquad
\bSigma_k = \frac{1}{t_{.k}} \sum_{i} t_{ik} (\by_i -\bmu_{k})(\by_i - \bmu_{k})^T.
\end{equation}
Putting everything together, we obtain the {\bf BEM} algorithm.

\paragraph{BEM algorithm:} Using the {\bf E} and {\bf M} steps defined above,
{\bf BEM} algorithm can be enumerated as follows:
\begin{description}
\item[Initialization] Set $\bt^{(0)},\br^{(0)}$ and $\btheta^{(0)}=(\bpi^{(0)},\brho^{(0)},\bbeta^{(0)},\bmu^{(0)}$, $\bSigma^{(0)}).$
\item[(a) Row-EStep] Compute $\bt^{(c+1)}$ using formula
\begin{equation}\label{eq:RowEStep}
t_{ik}^{(c+1)} =
\displaystyle\frac{\pi_k^{(c)} \displaystyle\prod_{jl}\left(f(x_{ij}|\by_i;\bbeta_{kl}^{(c)})
\phi(\by_i;\bmu_{k}^{(c)},\bSigma_{k}^{(c)})\right)^{r_{jl}^{(c)}} }
              {\sum_k \pi_k^{(c)} \displaystyle\prod_{jl}\left(f(x_{ij}|\by_i;\bbeta_{kl}^{(c)})\phi(\by_i;\bmu_{k}^{(c)},\bSigma_{k}^{(c)})\right)^{r_{jl}^{(c)}}}.
\end{equation}
\item[(b) Row-MStep] Compute $\bpi^{(c+1)}$, $\bmu^{(c+1)}$,  $\bSigma^{(c+1)}$ using equations (\ref{eq:prop-estimate}) and (\ref{eq:gauss-estimate}) and estimate
$\bbeta^{(c+1/2)}$ by solving maximization problem (\ref{eq:betakl-estimate}).
\item[(c) Col-EStep] Compute $\br^{(c+1)}$ using formula
\begin{equation}\label{eq:ColEStep}
r_{jl}^{(c+1)} = \frac{\rho_l^{(c)} \displaystyle\prod_{ik} f(x_{ij}|\by_i;\bbeta_{kl}^{(c+1/2)} )^{t_{ik}^{(c+1)}} }
              {\sum_l \rho_l^{(c)} \displaystyle\prod_{ik} f(x_{ij}|\by_i;\bbeta_{kl}^{(c+1/2)})^{t_{ik}^{(c+1)}}}.
\end{equation}
Observe that $r_{jl}$ does not depend of the density of $\by$.
\item[(d) Col-MStep] Compute $\brho^{(c+1)}$ using equations (\ref{eq:prop-estimate}) and estimate $\bbeta^{(c+1)}$ by solving maximization problem (\ref{eq:betakl-estimate}). 
\item[Iterate] Iterate {\bf (a)-(b)-(c)-(d)} until convergence.
\end{description}

\subsection{Selecting the number of blocks}
\label{sec:Selecting the number of blocks}
BIC is an information criterion defined as an asymptotic approximation of the logarithm
of the integrated likelihood (\cite{schwarz1978estimating}). The standard case leads to
write BIC as a penalised maximum likelihood:
$$
\mathrm{BIC} = -2\max_{\btheta} l(\btheta) + D \log(N)
$$
where $N$ is the number of statistical units and $D$ the number
of free parameters and $l(\btheta)$ defined in (\ref{eq:loglikelihood}).
Unfortunately, this approximation cannot be
used for LBM, due to the dependency structure of the observations
$(\bx,\by)$. However, a heuristic have been stated to define BIC
in \cite{keribin2012model} and \cite{keribin2015estimation}.
BIC-like approximations ICL lead to the following approximation
as $n$ and $m$ tend to infinity
\begin{equation}
\mathrm{BIC}(g,d) = -2\max_{\btheta}\log f(\bx, \by;\btheta) + (g-1)\log n  + \lambda\log n
+ (d-1)\log m + gd(p+1) \log(mn)
\end{equation}
with $\lambda$ the number of parameters of the $\by$ distribution.
For LBM, the intractable likelihood $f(\bx, \by;\btheta)$ is replaced by the maximized
free energy $\tilde{F}_C$ in (\ref{eq:fuzzycriterion}) obtained by the BEM algorithm.

\subsection{Measuring Influence of a Variable}
\label{sec:Measuring Influence of a Variable}
Let $j$ be fixed (a column of the matrix $\bx$). We would like to 
measure the effect of the variable
$\bx^j=(x_{ij})_{i=1}^n$ on $\by$. It is possible to obtain a measure
of this effect by looking to the posterior probability of $\by$.
\begin{lemma}
Let $(\bx,\bz,\bw)$ fixed. For $l=1,\ldots,d$ let $m_{l}$ denotes the number of columns
with label $l$, i.e $m_l = \#\{w_{jl} = 1,\; j=1,\ldots m\}$ and for a row $i$ fixed
let $m_{il}$ denotes the number of elements such that ${w_{jl} = 1}$ and $x_{ij} =1$,
i.e. $m_{il}=\#\{w_{jl}x_{ij} = 1,\; j=1,\ldots n\}$.
The posterior probability of the co-variable $\by$ is
\begin{align}\label{eq:co-posterior-row}
f(\by|\bx,\bz,\bw,\btheta) & \propto \prod_{i=1}^n \prod_{l=1}^{d} \pi_{z_i} \rho_{l}^{n_l}
  \logis(\by_i^T\bbeta_{z_i l})^{n_{il}}
  \left(1-\logis(\by_i^T\bbeta_{z_i l})\right)^{m_l-m_{il}}
  \phi(\by_i;\bmu_{z_{i}},\bSigma_{z_{i}}) \nonumber \\
  & \propto \prod_{i=1}^n  \pi_{z_i} \phi(\by_i;\bmu_{z_{i}},\bSigma_{z_{i}})
           \prod_{l=1}^{d} \rho_{l}^{n_l}   
  \frac{e^{n_{il}\by_i^T\bbeta_{z_i l}}}{\left(1+e^{\by_i^T\bbeta_{z_i l}} \right)^{m_l}}
\end{align}
Alternatively, for $k=1,\ldots,g$, let $n_k$ denotes the number of rows with label $k$,
i.e. $n_k = \#\{z_{ik}=1, i=1,\ldots,m\}$. The posterior probability of the co-variable
$\by$ is
\begin{equation}\label{eq:co-posterior-col}
f(\by|\bx,\bz,\bw,\btheta)  \propto \prod_{j=1}^m \rho_{w_{j}}
\prod_{k=1}^{g} \pi_{k}^{m_j} \prod_{i:z_i=k} \logis(\by_i^T\bbeta_{k w_j})^{x_{ij}}
  \left(1-\logis(\by_i^T\bbeta_{k w_j})\right)^{1-x_{ij}}
  \phi(\by_i;\bmu_{k},\bSigma_{k}).
\end{equation}
\end{lemma}
The proof of this lemma is straightforward and therefore omitted.

Assuming $\bz$ and $\bw$ known, we measure the influence of a variable 
using its contribution to the posterior probability. Fixing $j$,
taking the logarithm and eliminating terms independent of
$\bx^j$, we obtain the \textit{influence measure criteria}
\begin{align}
I(j)& = \log\rho_{w_{j}} +\sum_{i=1}^n {x_{ij}}\log\logis(\by_i^T\bbeta_{z_i  w_j})
 +\sum_{i=1}^n {(1-x_{ij})}\log\left(1-\logis(\by_i^T\bbeta_{z_i w_j})\right) \nonumber \\
 &  = \log\rho_{w_{j}} + \sum_{i=1}^n \left( x_{ij} \by_{i}^{T}\bbeta_{z_i w_j} - \log({1+\exp(\by_{i}^{T}.\bbeta_{z_i w_j})}) \right)
\end{align}
which is interpreted as the $\log$-contribution to the posterior distribution
(\ref{eq:co-posterior-col}) of the variable $\bx^j$.
Replacing the unknown labels $w_j$ and $z_i$ by their MAP estimators 
$\hat{w}_j$ and $\hat{z}_i$, we are able to sort the variables from the most
to the less influential.

\section{Examples}\label{sec:Applications}
\subsection{Simulated data}\label{subsec:SimulatedData}
\subsubsection{Computational time}\label{subsubsec:ComputationalTime}
\begin{figure}[H]
\begin{center}
\includegraphics[width=12cm]{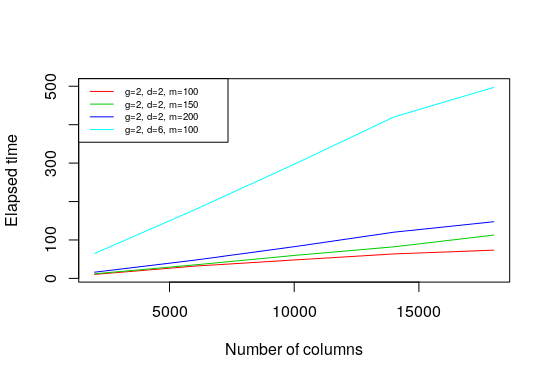}
\end{center}
\caption{computational elapsed time for $n=$ 2000, 6000, 10000, 14000 and 18000 (in minutes)
and for various values of $m$.}
\label{fig:time}
\end{figure}

We compute 80 times the elapsed time of the model for various configurations of the parameter
on a HP Zbook G3. The (averaged) computing time as a function of $m$ when $g=2$ for different values of $m$ (the number of columns) and when $d$ (the number of cluster in columns) take values 2 and 6 is plotted in figure \ref{fig:time} below
We can observe that as $n$ grows the elapsed time grows linearly, but that the slope increases
as $d$ (the number of class in columns) is increased.

\subsubsection{Error rate}\label{subsubsec:ErrorRate}
Next we simulate 80 times the number of columns well classified when $g=2$ and
for various configurations of $m$ and $d$. The cluster of a column is estimated
using the maximum a posterior (MAP) estimator
$$
\hat{w}_{j} = \arg\max_{l=1}^d r_{jl}.
$$

\begin{figure}[!htb]
 \centering
    \begin{minipage}{.5\textwidth}
        \centering
        \includegraphics[height=5cm]{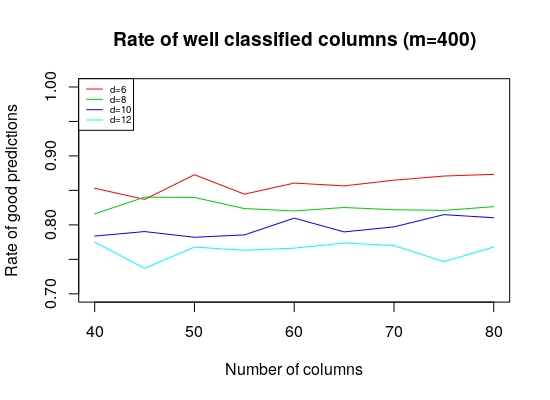}
\label{fig:prob1_6_2}
    \end{minipage}%
    \begin{minipage}{0.5\textwidth}
        \centering
        \includegraphics[height=5cm]{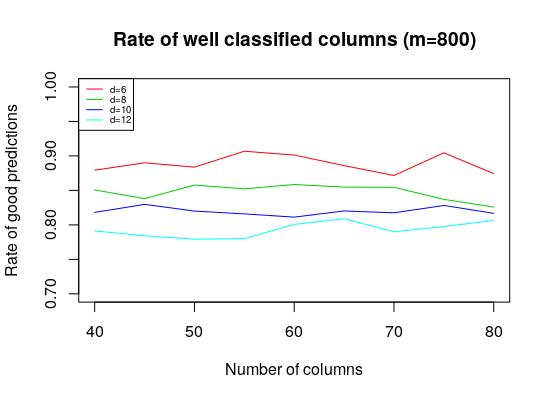}
        \label{fig:prob1_6_1}
    \end{minipage}
\caption{Rates of well classified columns when the number of rows is 400 and 800. The number of columns is between 40 and 80. The number of cluster is between 6 and 12. There is only two groups of rows.}
\end{figure}
From these partial results, we see that the number of bad classified columns labels increases as $d$ increases while it remains relatively constant with $m$. An other
salient feature is that when the number of individuals ($n$) is greater, this error
rate is lower. The number of well classified rows is stable near 0.9 for all tested
configurations of the parameters and is not displayed.

\subsection{Real  Data Analysis}\label{subsec:RealData}

\begin{figure}[htb]
 \centering
 \includegraphics[width=10cm, scale=4.0]{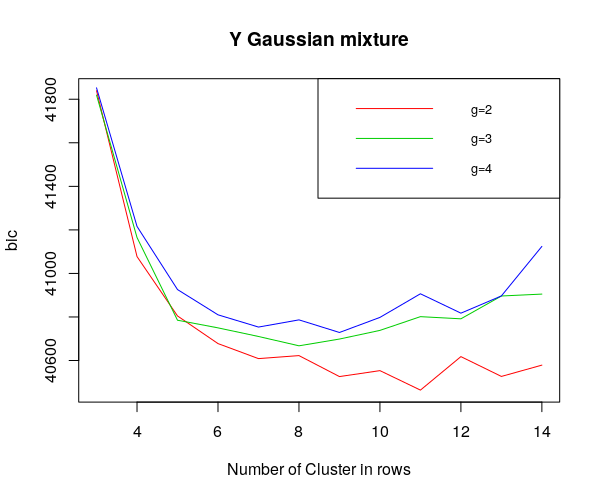}
 \caption{BIC computation for different values of $d$ and $g$. We observe that it
 is minimal for $g=2$ and $d=11$ among tested $d$ values $(1,\ldots,4)$ and $g$
 values $(2,3 4)$ .
 }
\label{fig:bic}
\end{figure}

Here, we study data from an epidemiological and genetic survey of malaria disease in Senegal.
Data were collected between 1990 to 2008. We worked on a dataset including $n=885$ individuals
with measured malaria risk score (phenotype) and genotype available on several candidate genes
for susceptibility/resistance to the disease. A total of $m=45$ Single Nucleotide Polymorphisms
(SNPs) was considered across these genes and was used as genetic variables. The malaria risk
score was a quantitative measure normally distributed  and was considered as a co-variable
for this co-clustering  method.
The SNPs are coded in dominant effect on the disease risk. 
Using the BIC criteria (see graph \ref{fig:bic}), we choose to focus on the model $d=2$ groups
of individuals and $g=11$ groups of SNPs.

\subsubsection{Analysis for phenotype data}
\label{subsubsec:MixtureAnalysis}
\begin{figure}[H]
\begin{minipage}{.49\linewidth}
\centering \includegraphics[width=6cm, height=60mm]{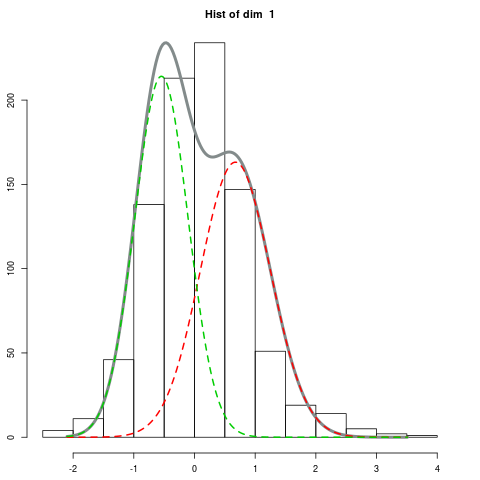}\\
\centering(a)
\end{minipage}
\begin{minipage}{.49\linewidth}
\centering  \includegraphics[width=8cm, height=60mm]{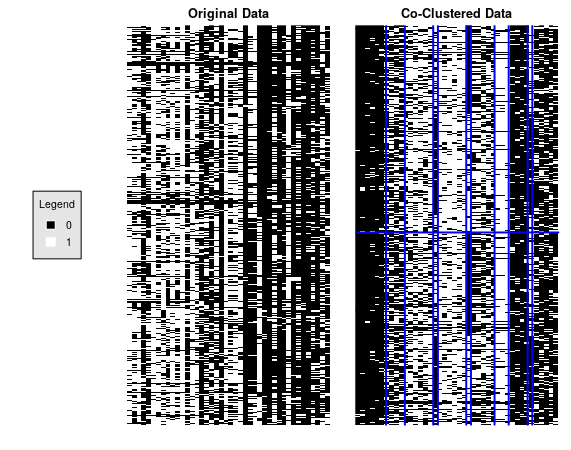}\\
\centering(b)
\end{minipage}
\caption{(a) 
- Empirical Distribution of the phenotype (histogram)
- Distribution of the susceptible (red)
- distribution of the resistant (green)
- mixing distribution (grey).\hspace{0.5cm}
(b) Array with the presence/absence of mutations before and after block-clustering}
\label{fig:data}
\end{figure}

The choice of a mixture model  or not depends on the application context.
In the case of genetic data, we are often interested in the comparison of the susceptible
and the resistant to a given phenotype.
In this application, we look for genes to explain the difference between susceptible
and resistant which justifies the use of a mixture model on the target variable.
After block-clustering, we find that the individuals are divided in two groups: 
the susceptibility category composed of a group of individuals with a value 
of phenotype essentially greater than zero and the resistant category composed of a group
of individuals with a value of phenotype essentially less than zero (see figure \ref{fig:data}).

Observe how the marginal distribution of the phenotype, which is uni-modal, becomes
multi-modal when conditioned by ($\bx,\bz$).

\subsubsection{Analysis for genotypes data }
\label{subsubsec:RealDataAnalysis}
We looked at the SNPs to determine which ones would potentially be involved
in malaria susceptibility / resistance.
\begin{figure}[H]
     \centering
     \includegraphics[width=10cm, scale=4.0]{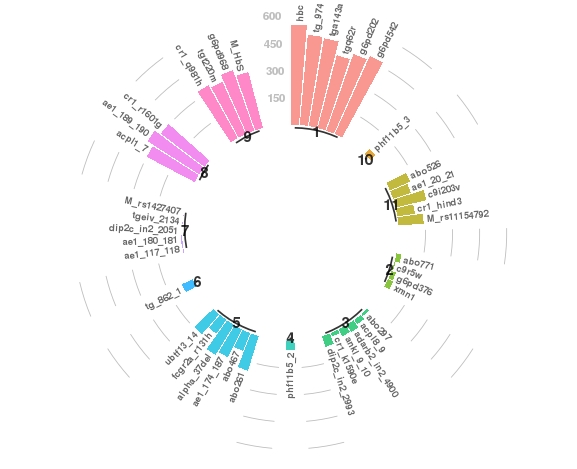}
     \caption{Representation of each block variable according to the influence measure }
     \label{fig:influence}
\end{figure}

The proposed methodology allowed the selection of the most significant SNPs according to the influence measure proposed in section \ref{sec:Measuring Influence of a Variable}.
The most frequent SNPs are grouped into the following classes: class 1 and 9. It is noted that the SNPs of these classes have been shown in the literature to have a high significance effect on malaria.
Most G6P and hemoglobin SNPs are grouped into these 2 classes.
Reviews from exiting literature gives us:
 Glucose-6-phosphate dehydrogenase (G6PD) deficiency is prevalent in sub-Saharan African populations and has been associated with protection against severe malaria \cite{maiga2014glucose, manjurano2015african, toure2012candidate, doi:10.1093/infdis/jir640}.
 Studies above haplotype analysis reveal that the G6PD locus is an under-balanced selection, suggesting a malaria protection mechanism based on modest frequency alleles and avoiding parasite attachment \cite{manjurano2015african}.
 Hemoglobins S and C (HbS and HbC respectively) are known to be two structurally variant forms of normal adult hemoglobin (HbA) resulting from distinct mutations in the $\beta$-globin gene.
 The protective effect of HbS against Plasmodium falciparum malaria has been shown by several authors 
\cite{beet1946sickle, allison1954protection, allison1954distribution}.
 In the case of HbC, the protection is highest in homozygous individuals with HbCC.
The proposed model confirmed the strong link between sickle cell polymorphism (HBS), blood group ABO (HBC) and falciparum malaria in the West African population.

 \subsubsection{Association between phenotype and genotypes }
 
\begin{figure}[H]
\centering
\includegraphics[width=10cm, scale=4.0]{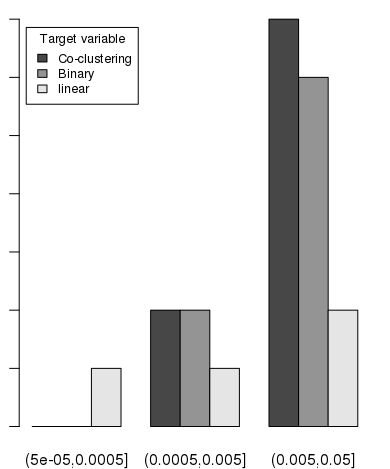}
\caption{Number of significant P-values for each method }
\label{fig:pvalue}
\end{figure}

The most common approach used in genetic data is the GWAS  method (Genome Wide Association
Studies). This method makes a linear regression of the quantitative phenotype on each
genotype variable.
By applying co-clustering with the phenotype as co-variable, we could obtain a dichotomy
of the phenotype. This dichotomy allows us to divide individuals into two categories:
susceptible and resistant. In this part, we compare the results of GWAS studies between 
the quantitative phenotype, the binary phenotype ($\un_{y_i\leq 0}$) and the (co-)clustered
phenotype. Figure \ref{fig:pvalue} shows that there are more signals at the 5\% threshold
for the  clustered phenotype compared to the two other phenotypes.
In summary the proposed methodology allows to detect more significant SNPs compared to the
quantitative and binary phenotype.

\section{Conclusion}\label{sec:conclusion}

In this article, our main contribution has been to develop a co-clustering model taking into
account a (mixture of) Gaussian co-variable. Applications have been made on simulated and real
data sets.
Our preliminary results are
confirmed in previous studies in Africa. The method offers good classification performance on
complex data sets (large number of variables and classes). This method can be useful
in a wide variety of classification problems with Gaussian predictors and will allow us to
discover new patterns of genes allowing to understand and evaluate the mechanism existing
between genetics and malaria in an African population particularly in a Senegalese rural area.
Further analysis could be done with more SNPs in another paper in preparation. 
Estimation is performed using a R package (with computational part in C++) that will be soon be available on the CRAN website \url{https://cran.r-project.org/}. Meanwhile the package is available on demand to the authors.

\bibliographystyle{ieeetr}
\bibliography{sample}

ieeetr
\appendix
\section{Computing the (rows and columns) E-Step}
\label{app:EStep}
For the E-Step $t_{ik}$ value maximize the fuzzy criterion
given in equation (\ref{eq:fuzzycriterion}). Derivative with
respect to $t_{ik}$ gives
$$
\frac{\partial\tilde{F}_C(\bt,\br;\btheta)}{\partial t_{ik}} = \log\pi_k + \sum_{j,\ell} r_{j\ell} \log f_{k\ell}(x_{ij},\by_i;\btheta) - \log t_{ik} - 1.
$$
Equating this equation to zero, taking exponential and recalling that
$\sum_k t_{ik} = 1$, we obtain that $t_{ik}$ is updated as  
$$t_{ik}^{(c+1)}=\frac{\pi_{k}^{(c)}  \prod_{j,l}  \left[  f(x_{ij},\by_i;\btheta^{(c)})\right]^{r_{jl}^{(c)}}  }{ \sum_{k}\prod_{j,l}  \left[  f(x_{ij},\by_i;\btheta^{(c)})\right]^{r_{jl}^{(c)}}}.
$$
For numerical reason, we prefer to compute the logarithm of this expression which is
$$\log(t_{ik}^{(c+1)}) \propto \log(\pi_{k}^{(c)}) + \sum_{j,l} r_{jl}^{(c)} \log f(x_{ij},\by_i;\btheta^{(c)} ). $$
Recall that (see equation \ref{eq:link})
\begin{eqnarray*}
   \log f(x_{ij}|\by_i;\bbeta_{kl}^{(c)}) & = & x_{ij}\log(\logis(\by_{i}^{T}\bbeta_{kl}^{(c)}))
+ 
(1-x_{ij})\log(1-\logis(\by_{i}^{T}\bbeta_{kl}^{(c)})) \\
   & = & \log(1-\logis(\by_{i}^{T}\bbeta_{kl}^{(c)})) + x_{ij}\log\left(\frac{\logis(\by_{i}^{T}\bbeta_{kl}^{(c)})}{1-\logis(\by_{i}^{T}.\bbeta_{kl})}\right) \\
   & = & \log(1+\exp(\by_{i}^{T}\bbeta_{kl}^{(c)})) + x_{ij} \by_i^T\bbeta_{kl}^{(c)}
\end{eqnarray*}
giving
$$
\log t_{ik}^{(c+1)} \propto \log \pi_{k}^{(c)} + \sum_{j,l} r_{jl}^{(c)} x_{ij} \by_i^T.\bbeta_{kl}^{(c)}
- \sum_{l} r_{.l}^{(c)} \log(1+e^{\by_i^T.\bbeta_{kl}^{(c)}}) + m\;\log\phi(\by_i;\bmu_k^{(c)},\bSigma_k^{(c)}).
$$
Similar computation gives for $r_{jl}$
$$
\log(r_{jl}^{(c+1)}) \propto \log\left(\rho_{l}^{(c)}\right)
+ \sum_{i,k} t_{ik}^{(c+1)} \left( x_{ij} \by_i^T\bbeta_{kl}^{(c+1/2)} - 
\log\left(1+e^{\by_i^T.\bbeta_{kl}^{(c+1/2)}}\right) \right).
$$
Observe that the Gaussian distribution does not depend of $j$ nor $l$.
This term become constant when summing over $i$ and $k$ and disappears
when $r_{jl}$ values are normalized.

\section{Computing the M-Step}
\label{app:MStep}
For the M-Step, we use a Newton-Raphson algorithm in order to solve the equation (\ref{eq:betakl-estimate}). For each pair $(k,l)$ the function to maximize 
can be written
$$
\ell_{k,l}(\bbeta)=\sum_{i,j}\left( r_{jl}t_{ik}x_{ij} \by_{i}^{T}\bbeta- r_{jl}t_{ik}\log({1+\exp(\by_{i}^{T}.\bbeta)})\right) 
$$
The first derivative with respect to the d-th coordinate $\beta_d$ is
$$
\frac{\partial \ell_{k,l}(\beta)}{\partial \beta_{d}}= 
\sum_{i,j}\left( r_{jl}t_{ik}x_{ij}y_{i,d}- r_{jl}t_{ik} y_{i,d} \frac{\exp(\by_{i}^{T}\bbeta)}{1+\exp(\by_{i}^{T}\bbeta)}\right)
$$
giving the following expression for the gradient
$$
\nabla_{\beta}\ell_{k,l}(\bbeta)=Y^{T}D(X-\bmu)
$$
with $Y=\left[\by_{i}\right]_{i=1}^{N}$,
$X=\left[ \sum_{j} r_{jl}x_{ij} \right]_{i=1}^{N} $, 
$\bmu=\left[ r_{.l}\frac{\exp(\by_{i}^{T}.\bbeta)}{1+\exp(\by_{i}^{T}\bbeta)} \right]_{i=1}^{N} $, $D=\diag(t_{ik})_{i=1}^{N}$
The second derivative with respect to $\beta_{d}$ and $\beta_{d^{'}}$ is
$$
\frac{\partial^{2} \ell_{k,l}(\bbeta)}{\partial \beta_{d}\partial \beta_{d^{'}}}= -\sum_{i,j}\left( r_{jl}t_{ik}y_{i,d}y_{i,d^{'}}\frac{\exp(\by_{i}^{T}\bbeta)}{(1+\exp(\by_{i}^{T}\bbeta))^{2}}\right) 
$$
giving the following expression for the hessian
$$
H_{\beta}=-Y^{t}DW Y \qquad \mbox{ with }\qquad
W=\diag\left( \frac{r_{.l}\,\exp(\by_{i}^{T}.\bbeta)}{(1+\exp(\by_{i}^{T}\bbeta))^{2}}\right)=\diag\left( r_{.l}\, \mu_i (1-\mu_i)\right)
$$

\end{document}